\documentclass[twocolumn]{aastex62}
\pdfoutput=1 
\usepackage{amsmath,amstext}
\usepackage[T1]{fontenc}
\usepackage{apjfonts} 
\usepackage[figure,figure*]{hypcap}
\usepackage{ulem}

\newcommand{\bsq}{{{b^2}}}
\newcommand{\dF}{{^{^*}\!\!F}}

\newcommand \RLC   {\ensuremath{ R_{\rm LC} }}
\newcommand \rlc   {\ensuremath{ \RLC } }
\newcommand \rco  {\ensuremath{ r_{\rm co} }}

\shorttitle{A Systematic Study of Accretion onto Oblique Pulsars} 
\shortauthors{Murguia-Berthier et al.}

\begin{document}
\title{From Feast to Famine: A Systematic Study of Accretion onto Oblique Pulsars with 3D GRMHD Simulations}

\author[0000-0003-2333-6116]{Ariadna~Murguia-Berthier}
\altaffiliation{NASA Hubble Fellow}
\affiliation{Center for Interdisciplinary Exploration \& Research in Astrophysics (CIERA), Physics \& Astronomy, Northwestern University, Evanston, IL 60202, USA}

\author[0000-0001-6173-0099]{Kyle Parfrey}
\affiliation{Princeton Plasma Physics Laboratory, Princeton, NJ 08540, USA}

\author[0000-0002-9182-2047]{Alexander Tchekhovskoy}
\affiliation{Center for Interdisciplinary Exploration \& Research in Astrophysics (CIERA), Physics \& Astronomy, Northwestern University, Evanston, IL 60202, USA}

\author[0000-0003-2982-0005]{Jonatan Jacquemin-Ide}
\affiliation{Center for Interdisciplinary Exploration \& Research in Astrophysics (CIERA), Physics \& Astronomy, Northwestern University, Evanston, IL 60202, USA}

\begin{abstract}
Disk-fed accretion onto neutron stars can power a wide range
of astrophysical sources ranging from X-ray binaries, to accretion powered millisecond pulsars, ultra-luminous X-ray sources, and gamma-ray bursts. A crucial parameter controlling the gas--magnetosphere interaction is the strength of the stellar dipole. In addition, coherent X-ray pulsations in many neutron star systems indicate that the star's dipole moment is oblique relative to its rotation axis.
Therefore, it is critical to systematically explore the 2D parameter space of the star's magnetic field strength and obliquity, which is what this work does, for the first time, in the framework of 3D general-relativistic magnetohydrodynamics.
If the accretion disk carries its own vertical magnetic field, this introduces an additional factor: the relative polarity of the disk and stellar magnetic fields.
We find that depending on the strength of the stellar dipole and the star--disk relative polarity, the neutron star's jet power can either increase or decrease with increasing obliquity.
For weak dipole strength (equivalently, high accretion rate), the parallel polarity results in a positive correlation between jet power and obliquity, whereas the anti-parallel orientation displays the opposite trend. For stronger dipoles, the relative-polarity effect disappears, and jet power always decreases with increasing obliquity.
The influence of the relative polarity gradually disappears as obliquity increases. 
Highly oblique pulsars tend to have an increased magnetospheric radius, a lower mass accretion rate, and enter the propeller regime at lower magnetic moments than aligned stars. 

\end{abstract}


\section{Introduction}\label{section:intro}

Accretion onto neutron stars (NSs) can power a wide range of astrophysical sources ranging from X-ray binaries \citep{2015A&ARv..23....2W}, to accretion powered millisecond pulsars \citep{2009Sci...324.1411A}, ultra-luminous X-ray sources \citep{2014Natur.514..202B,2017Sci...355..817I},  magnetar-powered superluminous supernovae \citep{2010ApJ...719L.204W} and gamma-ray bursts \citep{1992ApJ...392L...9D,2011MNRAS.413.2031M}. The fact that NSs have a surface and show periodic variability provides a unique opportunity to turn them into astrophysical laboratories of accretion by directly measuring their spins and spin derivatives: this enables direct measurements of the accretion flow and magnetospheric torques experienced by the star, a luxury not afforded to us by black holes (BHs) where spin measurements are scarce and often come with significant uncertainties \citep{2006ARA&A..44...49R,2009ApJ...701L..83S,2014SSRv..183..277R}. However, these data-rich NS observations are \emph{interpretation-limited} due to the highly nonlinear gas-magnetosphere interactions that confound analytic models and the scarcity of first-principles 3D models that describe the interaction between the NS magnetosphere and infalling gas.

Simulating the gas-magnetosphere interactions in the NS context is indeed a challenging problem, because it involves enormous density and magnetization contrasts: light relativistically magnetized magnetospheric plasma interacting with heavy weakly magnetized infalling gas. The two regions require seemingly mutually exclusive approaches: a force-free approach capable of handling the high magnetization of the magnetosphere (but incapable of handling the gas) and a general-relativistic magnetohydrodynamical (GRMHD) approach capable of handling moderately magnetized fluid (but incapable of handling the extreme magnetization of the magnetosphere). 

Several groups have performed studies of gas-magnetosphere interactions in the NS context \citep{2017MNRAS.469.3656P,2017ApJ...851L..34P,2022MNRAS.515.3144D,2023arXiv231104291P,2023arXiv231105301D}. 
Recently, \citet{2023arXiv231104291P} self-consistently simulated in 3D a neutron star surrounded by an accretion disk using a hybrid GRMHD approach, which combines the advantages of the force-free and GRMHD techniques. They studied how the relative orientation of the large-scale NS and disk magnetic field vectors affected magnetized NS accretion, as a function of NS dipole strength. They found that similar to 2D simulations \citep{2017ApJ...851L..34P}, the reconnection of disk and magnetospheric magnetic fields can aid in opening closed stellar magnetic field lines and result in an increased NS spindown power, launching a relativistic jet. However, in contrast to 2D, even when the disk and magnetospheric fields are parallel to each other and cannot reconnect, the field line opening can still happen due to the \emph{interchange slingshot}, a version of the magnetic interchange instability. They also found that this mechanism only activates when the NS magnetic fields are strong enough to push the accretion flow away (i.e., in the propeller regime), or close to doing so. Therefore, both the field strength and its orientation, or the tilt of the NS magnetosphere, crucially affect the disk-magnetosphere interactions.

Indeed, the magnetic and rotational axes of accreting NSs are typically tilted relative to one other, with tilt angles sometimes reaching as high as $60^\circ$ \cite[e.g., XSS J12270-4859,][]{2014MNRAS.444.3004D}. How does such a large tilt affect the system?  Does it decrease the magnetospheric radius at constant mass accretion rates \citep{1997ApJ...475L.135W,2018A&A...617A.126B}? Does it increase the pulsar's spindown power similar to isolated pulsars \citep{2006ApJ...648L..51S,2013MNRAS.435L...1T,2016MNRAS.457.3384T}? If so, can this make it harder for gas to accrete and, possibly, even bring about the propeller state, in which the  disk--magnetosphere interaction expels the accretion flow?

Due to the lack of symmetries and the non-linearity of gas--magnetosphere interactions, direct multidimensional simulations provide an attractive line of attack on this critical problem.  \citet{2003ApJ...595.1009R,2004ApJ...610..920R,2013MNRAS.430..699R,2021MNRAS.506..372R} explored the interaction between a non-relativistic oblique dipole and an accretion flow via an $\alpha-$prescription, and \citet{2012MNRAS.421...63R} via the magnetorotational instability (MRI, \citealt{1991ApJ...376..214B,1991ApJ...376..223H,1998RvMP...70....1B}).
They found that obliquity dramatically affects the geometry of the incoming streams, mass accretion rate, neutron star spindown power, and the radiative signatures from the hotspots produced by the infalling streams on the stellar surface.

In this paper, we perform the first systematic exploration of magnetized accretion onto a rapidly rotating oblique NS dipolar magnetosphere, for a wide range of obliquities and stellar dipolar strengths, using the 3D hybrid-GRMHD approach. This enables us to self-consistently handle relativistically magnetized magnetospheric plasma alongside dense weakly magnetized non-relativistic gas, and to resolve the MRI. In Sec.~\ref{section:num}, we discuss the problem setup and numerical method, in Sec.~\ref{section:results} we investigate the effects of the obliquity angle (Sec.~\ref{sec:obliquity}) and stellar magnetic dipole moment (Sec.~\ref{sec:mu}), and finish with the special case of a perpendicular rotator (Sec.~\ref{sec:perp-rot}). We conclude in Sec.~\ref{sec:summary}.

\section{Numerical Method and Problem Setup}\label{section:num}

We solve the equations of GRMHD using a modified version of the \textsc{harmpi} code \citep{2019ascl.soft12014T}, which is based on the \textsc{harm} code \citep{2003ApJ...589..444G,2006ApJ...641..626N}:
\begin{align}
      \nabla_\mu (\rho u^\mu)&=0, \\
      \nabla_\mu T^{\mu}_{\nu}&=0, \\
    \nabla_\mu \dF^{\mu\nu}&=0,
\end{align}
where $\rho$ is the fluid-frame mass density, $u^\mu$ is the  4-velocity, and $\dF^{\mu \nu}=b^{\mu}u^{\nu}-b^{\nu}u^{\mu}$ is the dual of the electromagnetic tensor. We set $G = M_* = c = 1$ and work in Lorentz-Heaviside units such that the fluid-frame magnetic pressure is $b^2/2$, where  $b^\mu = \dF^{\nu \mu} u_\nu$ 
is the fluid-frame magnetic field 4-vector. The stress energy tensor takes the form:
\begin{equation}
    T^{\mu\nu}= \rho h u^\mu u^\nu + P g^{\mu\nu}+\bsq u^\mu u^\nu + \frac{1}{2} \bsq g^{\mu\nu} - b^\mu b^\nu \;,
\end{equation}
where $h=\left(1 + \epsilon + P/\rho \right)$ is the specific enthalpy, $P$ is the gas pressure, $\epsilon$ is the specific internal energy density,  and $g_{\mu \nu}$  is the metric tensor.

\begin{figure*}[t!]
\includegraphics[scale=0.39]{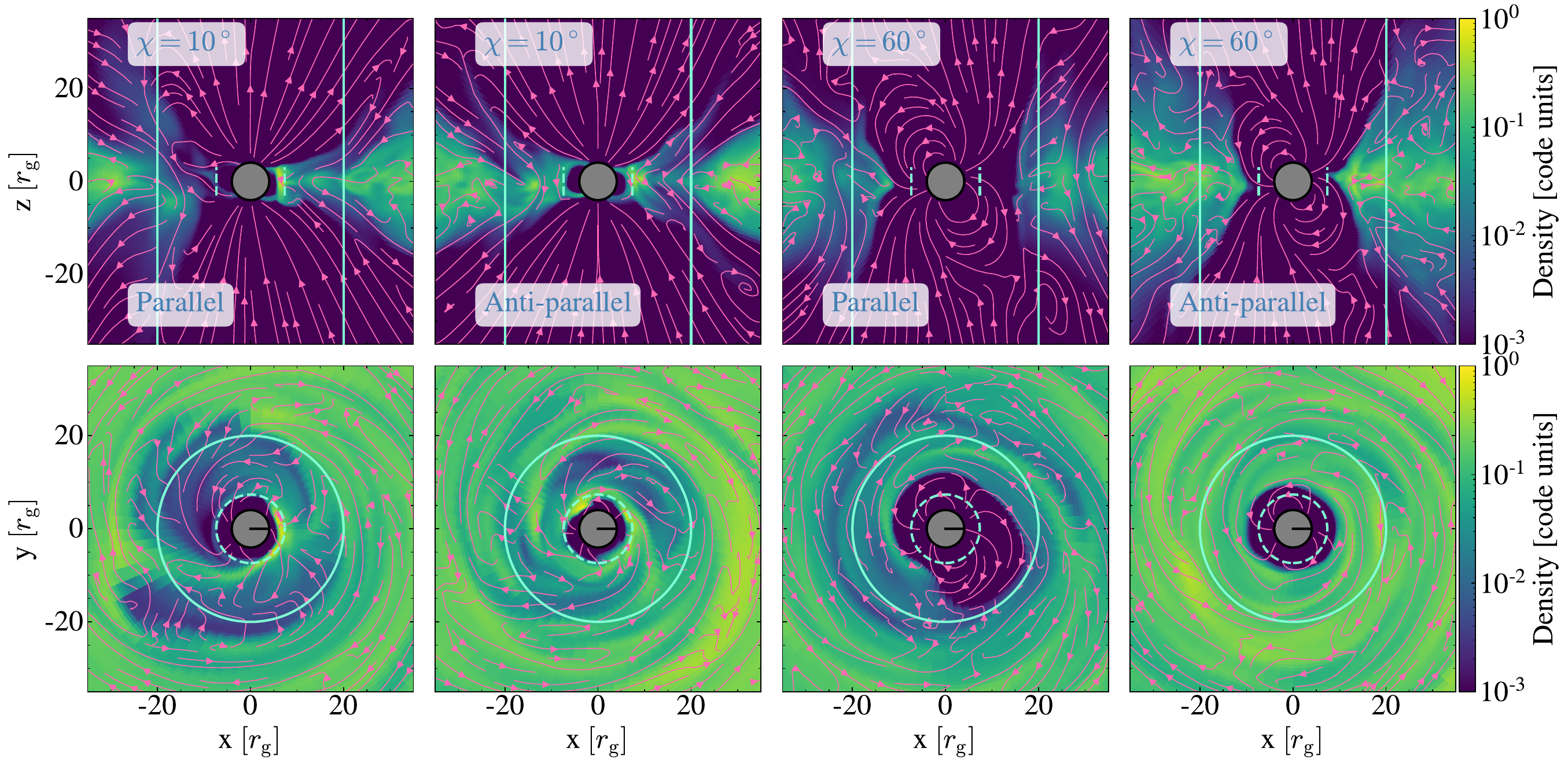}
\caption{Density slices through the  $\mu-\Omega$ (\textit{top} row) and equatorial (\textit{bottom} row)  planes of our simulations with a magnetic moment of $\mu=20$ reveal that the increase in magnetic obliquity brings the system closer to the propeller state. We show the system at $t=14{,}200\, r_{\rm g}/c\approx 113\,P_*$, 
for both parallel and anti-parallel configurations. {\bf [Left 2 panels]:} At low obliquity, $\chi = 10^\circ$, the gas reaches the NS. {\bf [Right 2 panels]:} However, at high obliquity, $\chi = 60^\circ$, the magnetosphere pushes most of the gas outside of the corotation radius, shown with the dashed lines. As a result, the system enters the propeller regime, and the gas struggles to reach the NS.  Directed pink lines trace out the magnetic field in the image plane. The solid cyan lines indicate the light cylinder.}
\label{fig:ct_mu}
\end{figure*}

\begin{figure}[t!]
\includegraphics[scale=0.54]{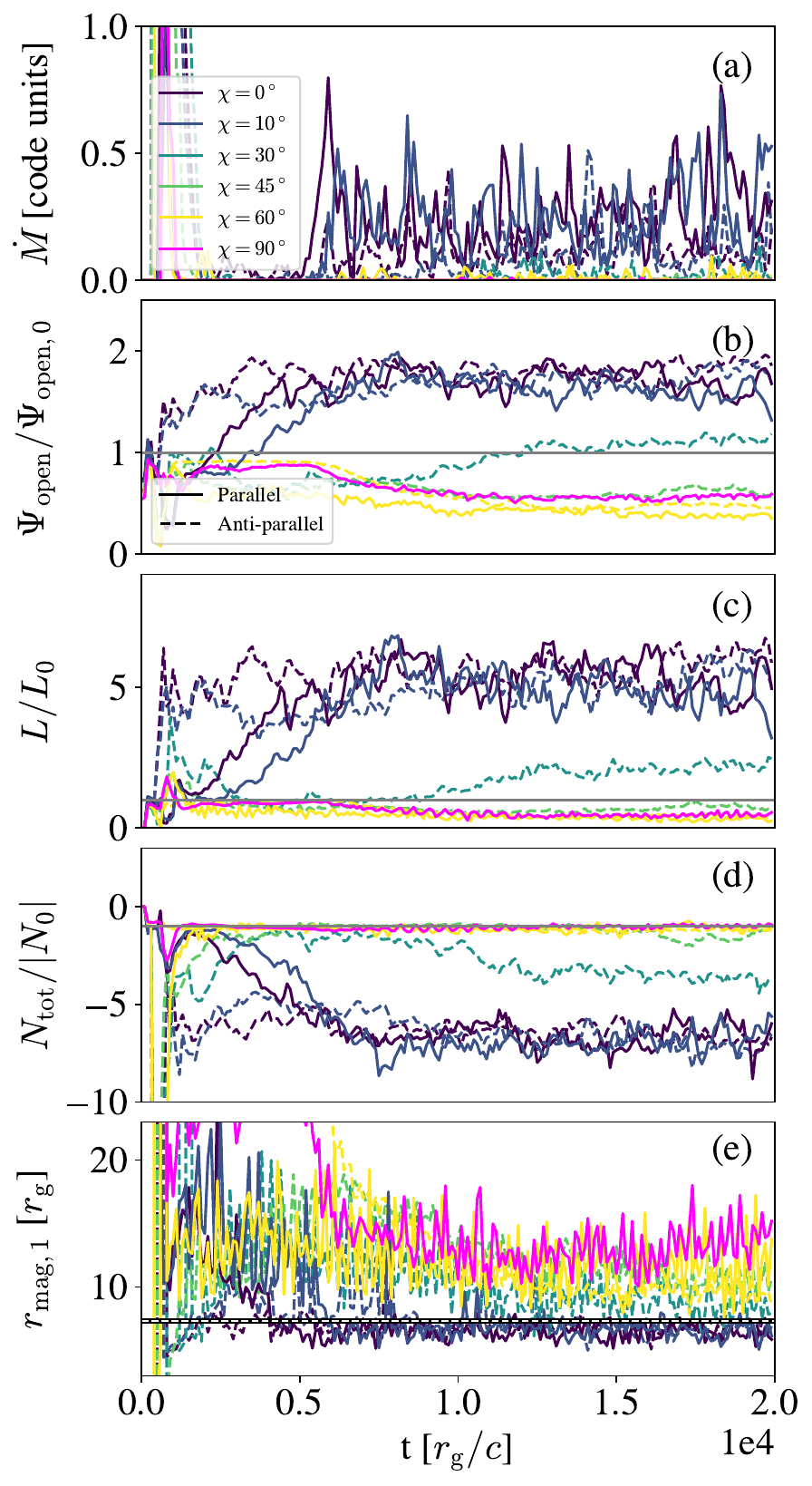}
\centering
\caption{ Different quantities as a function of time for $\mu=20$. The dashed lines correspond to simulations in which the magnetic field of the torus and the NS are anti-parallel, and the solid line represents simulations in which they are parallel. The magenta line represents the perpendicular simulation. From \textit{top} to \textit{bottom}, the quantities are: (a) mass accretion rate  at the NS surface; (b) open magnetic flux at $r=\rlc$ normalized to an isolated pulsar; (c) jet power at $r=\rlc$, similarly normalized; (d) total torque, similarly normalized; (e)  magnetospheric radius. See section~\ref{section:results} for a detailed explanation of the quantities and how they were calculated. The horizontal lines represent $\Psi_{\rm open}=\Psi_{\rm open,0}$, $L=L_0$, $N_{\rm tot}=-|N_0|$, $r_{\rm mag}=\rco$.}
\label{fig:quantities_time}
\end{figure}

\begin{figure*}[t!]
\includegraphics[scale=0.39]{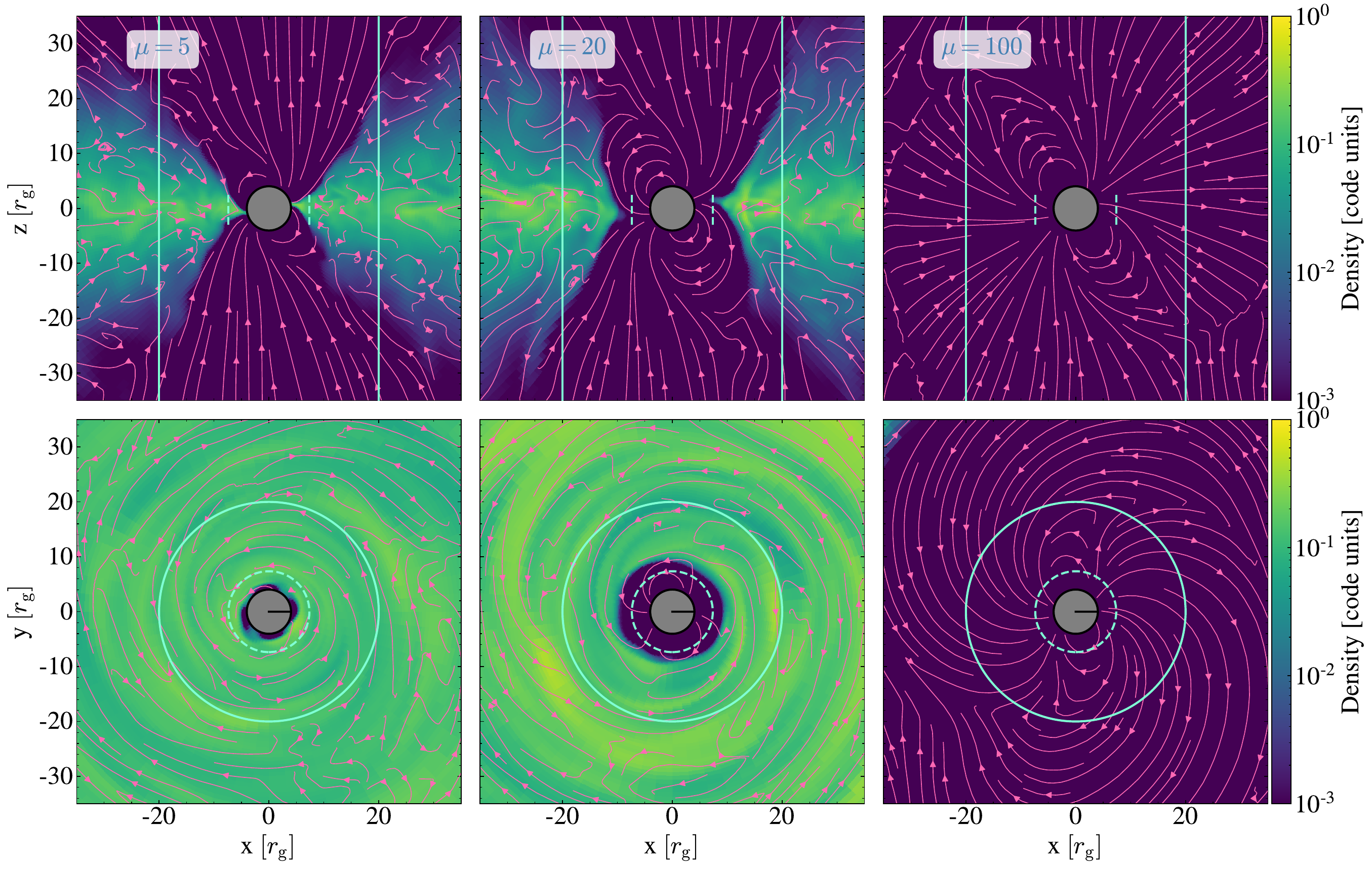}
\caption{Density snapshots of the  $\mu-\Omega$ plane (\textit{top} panel) and the equatorial plane (\textit{bottom} panel) for simulations with $\chi=60^\circ$, in an anti-parallel configuration. We show different magnetic moments at time $t=14{,}200\,r_{\rm g}/c$, or around 113 NS spin periods. The pink lines represent the in-plane magnetic field component. The solid cyan line is placed at the light cylinder, and the dashed cyan line is the corotation radius.}
\label{fig:diff_mu}
\end{figure*}

\begin{figure*}[t!]
\includegraphics[scale=0.39]{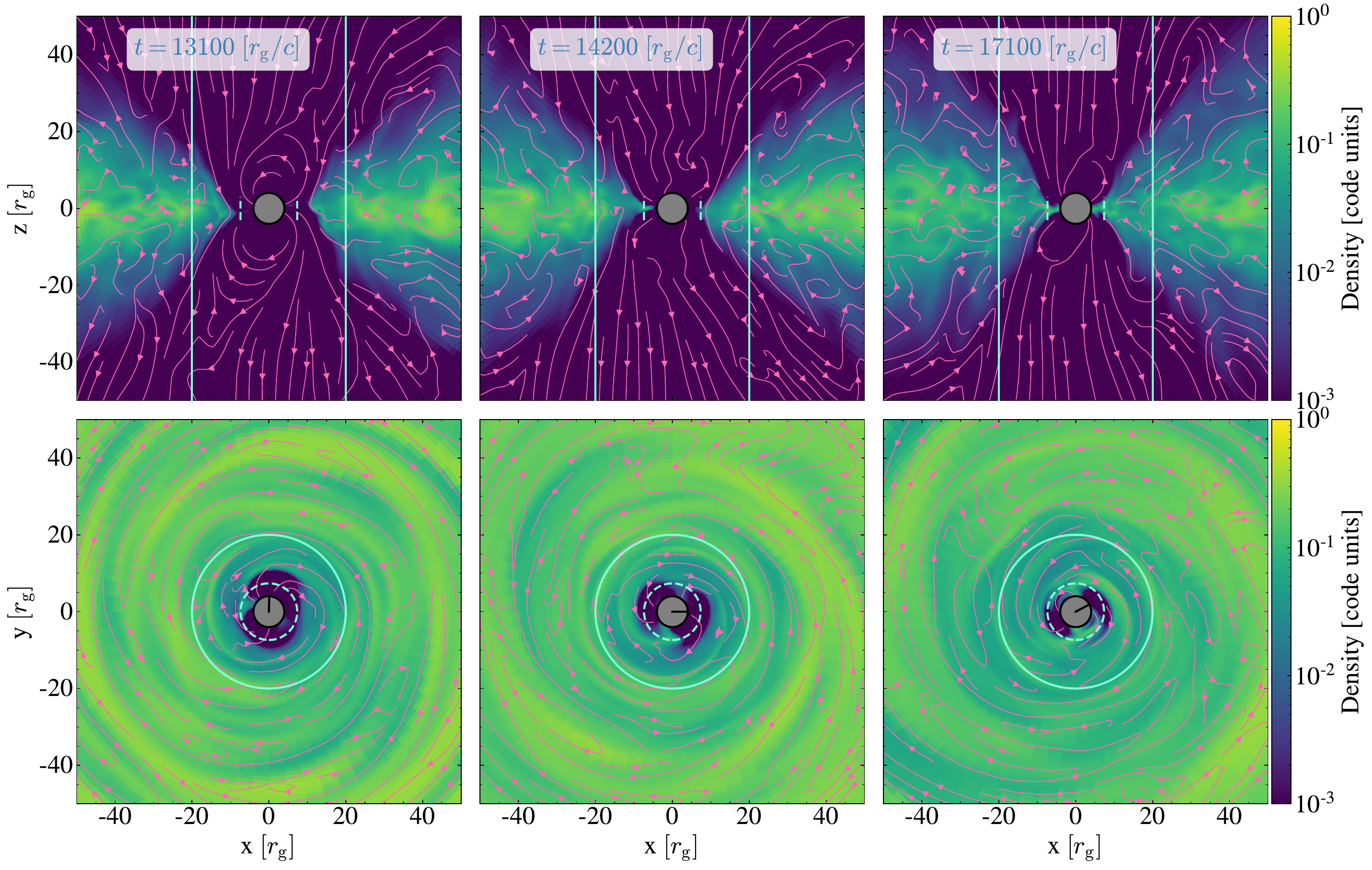}
\caption{Density snapshots of a simulation with a magnetic moment of $\mu=10$ at different times, where the star's magnetic moment is perpendicular to its rotation axis. \textit{Top} row panels show different planes, from left to right: (a) the plane perpendicular to $\mu-\Omega$, (b) the $\mu-\Omega$ plane, and (c) a plane between those two. The \textit{bottom} panel is the equatorial plane.  The pink lines represent the magnetic field of the in-plane component. The solid cyan line is placed at the light cylinder, and the dashed cyan line is the corotation radius.}
\label{fig:perpendicular}
\end{figure*}

We place the NS stellar surface at $r_*=4\, r_{\rm g}$, where $r_{\rm g}=G M_* / c^2$ is the gravitational radius of the star. We adopt the angular velocity of the NS to be $\Omega_*=0.05\, c/r_{\rm g}$, which places the light cylinder at $\rlc\equiv c/\Omega_*=20\, r_{\rm g}$ and sets the NS rotational period, $P_* = 2\pi/\Omega_*$. We use the Kerr metric, with a dimensionless spin of $a=1/3$.

The Keplerian angular velocity, $\Omega_{\rm K}(r) = 1/(r\sqrt{r} + a)$, determines the corotation radius, $\rco  \approx 7.3\, r_{\rm g}$, at which where the NS spin and Keplerian angular velocities are the same. We add a dipolar magnetic field by setting the magnetic vector potential \citep{1983ApJ...265.1036W}:
\begin{align}
    A_{\phi, \rm WS}(r,\theta,\phi) &= \Phi\sin{\theta}(\cos{\chi}\sin{\theta}-\sin{\chi}\cos{\theta}\cos{\phi}) ,\\
    A_{\theta, \rm WS}(r,\theta,\phi) &= -\Phi\sin{\phi}\sin{\chi},
\end{align}
where
\begin{equation}
    \Phi=\frac{3 \mu \sin^2\theta}{2M_*} \left[ \lambda^2 \ln(1-\lambda^{-1}) + \lambda + \frac12\right],
    \label{eq:WassShap}
\end{equation}
$\lambda = r/2 M_*$, and $\mu$ is the magnetic moment of the star in units of $4\pi r_{\rm g}^3 \sqrt{\rho_{\rm max}c^2}$. %

For the rotating magnetized NS boundary conditions, we set the surface electric fields in our constrained transport scheme such that at each time step they result in the surface radial magnetic field, which is a discretization of the initial dipolar magnetic flux rotated by an angle $\Delta\phi = \Omega_* t$, where $t$ is the current simulation time  \citep{2013MNRAS.435L...1T,2014MNRAS.441.1879P}.

We include an equilibrium hydrodynamic torus \citep{1976ApJ...207..962F,1985ApJ...288....1C}, with the inner radius at $r_{\rm in} = 40r_{\rm g}$ and the radius of maximum pressure at $r_{\rm max} = 60r_{\rm g}$. Inside the torus, we add a single poloidal (i.e., in the $R$- and $z$-directions) magnetic field loop that follows the lines of constant density, described by the magnetic vector potential, $A_{\phi,\rm disk} \propto \rho^2 r^{5} $. %
We normalize the disk magnetic field so that the ratio of the gas to magnetic pressure maxima is $\max P/ \max(b^2/2)=100$. %
Note that we have the freedom to orient the disk magnetic field loop to be either parallel or anti-parallel to the dipolar magnetic field of the NS, and in this work we will consider both magnetic field orientations. We note that, unlike \citet{2023arXiv231104291P}, in these simulations we do not deform the magnetic field around the torus. This results in accretion occurring earlier and a different profile of vertical magnetic flux being dragged inward by the disk over time.  %

We adopt the approach of handling both MHD and force-free regions in a single simulation in a self-consistent way \citep{2017ApJ...851L..34P,2023arXiv231104291P}. Namely, we introduce a passive scalar, $\mathcal{F}$, that we evolve as
\begin{equation}
    \nabla_\mu (\mathcal{F} \rho u^\mu) = 0.
\end{equation}
The scalar enables the algorithm to distinguish the magnetospheric and gaseous regions inside the light sphere $r\le R_{\rm LC}$: $\mathcal{F}\approx 1$ in the magnetospheric, or force-free, regions, and $\mathcal{F}\approx 0$ where the accreting gas dominates. When $\mathcal{F} = 0$ the GRMHD equations are used without modification, while when $\mathcal{F} = 1$ the non-force-free degrees of freedom are driven towards force-free-like values over a characteristic timescale; intermediate $\mathcal{F}$ values smoothly interpolate between these extremes. The GRMHD evolution is unmodified  in all cells for $r>R_{\rm LC}$.  %

We performed a suite of simulations covering a wide range of dipolar magnetic field strengths, $\mu$, and obliquity angles, $\chi$, where $\chi=0^\circ$ denotes an aligned rotator. Namely, we performed simulations for $\mu=5,10,20,100$, each for several obliquity angles, $\chi=0^\circ,10^\circ,60^\circ,90^\circ$, and for both the anti-parallel and parallel configurations of torus magnetic field.\footnote{Note that the two torus magnetic field configurations are equivalent for $\chi = 90^\circ$.} We also simulated the intermediate, $\chi=30^\circ,45^\circ$, cases for $\mu=20$ and anti-parallel torus magnetic field. 

For comparison, we repeated all of the above models for an isolated pulsar (i.e., without the initial torus). Our simulations have a resolution of  $N_{ r}=192$, $N_\theta=128$, $N_\phi=64$. We evolve all simulations until $t_{\rm F} = 20{,}000\, r_{\rm g}/c \approx 150P_*$.

\section{Results}\label{section:results}

\subsection{Effect of obliquity angle}
\label{sec:obliquity}

Figure~\ref{fig:ct_mu} shows the vertical, $\mu-\Omega$, and equatorial, $x-y$, slices through the density in our $\mu=20$ simulations for low, $\chi=10^\circ$, and high, $\chi = 60^\circ$, obliquities, at $t = 14200r_{\rm g}/c \approx 113P_*$. Directed pink lines show the image plane magnetic field.

At the beginning of the simulation, magnetic stresses are induced in the torus by its differential rotation, triggering the MRI.  As angular momentum is transported outwards, material from the torus is transported inwards, onto the NS. 
When the obliquity angle is low, the pressure due to the star's closed magnetic field lines halts the accretion flow and redirects some of the material to the poles. The radius at which this happens is known as the magnetospheric radius. If the magnetic field the torus presents to the NS is anti-parallel to the NS closed dipolar field, the two fields can reconnect: this will open NS field lines that would be closed in an otherwise isolated pulsar. The additional open field lines add to the pulsar wind, and create a Poynting-flux-dominated relativistic jet. This can be seen on the panels for $\chi=10^\circ$ (low obliquity), for the anti-parallel configuration, in the $\mu-\Omega$ plane. We see ordered magnetic field lines forming what in 3D is a helical structure. We can also see how material from the torus is accreted onto the NS in a ``ring''-like manner. 
On the other hand, if the magnetic field lines of the torus and the dipolar field are parallel, reconnection is not possible, but, for high enough magnetic moment (like $\mu=20$), magnetic field lines can open due to the interchange slingshot instability \citep{2023arXiv231104291P}.

If we increase the obliquity angle, instead of a ``ring'', material will be accreted onto the NS via  a single, thin stream onto each magnetic pole. This is illustrated in the $\mu-\Omega$ panels of the  $\chi=60^\circ$ simulation in Figure~\ref{fig:ct_mu}. Although it is not seen in the shown snapshots, at several times, and at several different planes, the thin stream will fall onto the NS and be accreted. Similar results were found in non-relativistic MHD simulations by \citet{2003ApJ...595.1009R,2004ApJ...610..920R,2021MNRAS.506..372R}. Even though the material from the torus has an easier, natural path to the NS, the stream is thinner than the ``ring''-columns, and less material is accreted onto the NS. 

Aside from the different accretion stream geometry, there is another interesting effect. Since the equatorial region of the NS magnetic field is no longer in the same plane as the material being accreted, magnetic reconnection should occur less frequently. This results in more closed lines. Since there is less additional open flux to strengthen the pulsar wind, this effect may severely inhibit jet formation.

In Figure~\ref{fig:quantities_time}, we plot key quantities as a function of time. From  \textit{top} to \textit{bottom} panels, we show the mass accretion rate, open magnetic flux, jet power, NS torque, and magnetospheric radius. We calculate the quantities as in \citet{2023arXiv231104291P}, and normalize some of them (open flux, jet power, and torque) to those of an isolated pulsar at the same obliquity angle, obtained from simulations without a torus and indicated with a ``0'' subscript.

The mass accretion rate is calculated by integrating the flux of mass at the NS surface. The open magnetic flux, calculated at the light sphere, $r=\rlc$, is found by integrating over the force-free ($\sigma>1$) region, 
\begin{equation}
\Psi_{\rm open} = \int_0^{2\pi}\int_0^\pi |B^r(r,\theta,\phi)| \sqrt{-g}\, {\rm d}\theta\, {\rm d}\phi,
\end{equation}
where $\sigma = b^2/\rho h$ is the magnetization parameter and $B^r$ is a measure of the radial magnetic field. %

The jet power, also calculated at the light sphere and for the force-free region, is measured as
\begin{equation}
    L = - \int_0^{2\pi}\int_0^\pi  \left(b^2 u^r u_t - b^r b_t\right) \sqrt{-g}\, {\rm d}\theta\, {\rm d}\phi.
\end{equation}

The torque is calculated at the NS surface, and obtained by integrating the hydrodynamic and electromagnetic contributions:
 \begin{equation}
     N_{\rm tot}= -\int_0^{2\pi}\int_0^\pi   \left( b^2 u^r u_\phi - b^r  b_\phi   +\rho h u^r u_\phi \right) \sqrt{-g}\, {\rm d}\theta\, {\rm d}\phi.
 \end{equation}
In this convention, a negative torque results in spindown of the star. 

We define the magnetospheric radius as the location at which $\sigma=1$ in the equatorial plane. Due to the geometry of the problem, we calculate two different magnetospheric radii: $r_{\rm mag, 1}$ is measured in the $\mu-\Omega$ plane, while $r_{\rm mag, 2}$ is defined in the plane perpendicular to the $\mu-\Omega$ plane. 

Figure~\ref{fig:quantities_time} shows these quantities as a function of time for  the simulations performed with $\mu=20$. As the simulation starts, the MRI in the accretion disk transports the angular momentum outward and results in the infall of the torus material towards the NS. At $t\approx7{,}500\, r_{\rm g}/c$, the material reaches the surface of the NS, and at roughly $t = 10{,}000\, r_{\rm g}/c$ the quantities settle into an approximate steady state. 

If we focus first on the magnetospheric radius (Figure~\ref{fig:quantities_time}e), it increases as we increase obliquity, eventually becoming larger than the corotation radius. When $r_{\rm mag} \gtrsim \rco$, the pulsar enters the so-called propeller regime, and little to no material is accreted as it would have to overcome the centrifugal barrier.  

High-obliquity pulsars tend to reach the propeller stage at lower magnetic moments than their non-oblique counterparts. As seen in Figure~\ref{fig:ct_mu}, in the $\mu-\Omega$ plane of the $\chi=60^\circ$ pulsars, the interactions with the magnetosphere redirect the material from the torus onto each of the star's magnetic poles through a single, thin stream. 

The magnetic field strength of a dipole is twice as high at its pole as at its equator, and therefore tilting the pole increases the field strength at the rotational equator in the $\mu$--$\Omega$ plane. This may partly explain why the magnetospheric radius in the $\mu$--$\Omega$ plane ($r_{\rm mag, 1}$) is consistently larger than the measurement in the perpendicular plane ($r_{\rm mag, 2}$), where the field strength is unchanged from its aligned-rotator value.

From Figure~\ref{fig:quantities_time}c, which shows the jet power, we can see that as we increase obliquity  the jet power decreases. At some angle (around $\chi=45^\circ$), it resembles the isolated-pulsar value, and as we further increase the obliquity angle the jet power becomes less than the isolated-pulsar spindown power. The open magnetic flux in Figure~\ref{fig:quantities_time}b has a similar behaviour. Both of these effects are partly due to the increase in the magnetospheric radius. At high obliquity angles, the magnetospheric radius is larger than the corotation radius the majority of the time. Material now has to break the centrifugal barrier in order to accrete, thus, the mass accretion rate will be inhibited (as can be seen in the mass accretion rate panel), and less magnetic flux will be open.

We also note, when comparing different obliquity angles at the same magnetic moment, that the isolated-pulsar spindown power is larger at higher obliquity. \citet{2006ApJ...648L..51S} found a difference of two between aligned and orthogonal rotators (the difference may be slightly higher; \citealt{2013MNRAS.435L...1T}). Since in the isolated case $L\propto\mu^2$, the factor of two translates into a difference in the effective magnetic moment of $\sqrt{2}$. In our simulations, even though at $\mu=5$ all our runs are in the accretor (non-propeller) regime, at $\mu=10$ and 20 we have the interesting case where low-obliquity pulsars are in the accretor regime, whereas at larger obliquity angles the pulsar is already in the propeller regime.

\subsection{Effect of the magnetic moment and star--torus relative field orientation}
\label{sec:mu}

In Figure~\ref{fig:diff_mu}, we show simulations of a $\chi=60^\circ$, anti-parallel magnetic field configuration with different magnetic moments. At lower magnetic moments ($\mu=5$), we still see the formation of a single, thin  accretion stream in the $\mu-\Omega$ plane.

As we increase the magnetic moment, the pressure from the magnetosphere increases, and the stream becomes thinner. There is very little dense material inside the corotation radius, and thus the mass accretion rate drops. Notice that at intermediate magnetic moment ($\mu=20$), there is still some accretion, although it is severely inhibited. At even higher magnetic moment ($\mu=100$), the magnetospheric pressure is high enough that there is no accretion onto the NS and the pulsar wind dominates. In our simulations at $\mu=100$ the pulsars are fully in the propeller regime for all obliquity angles.

As pointed out in \citet{2017ApJ...851L..34P,2023arXiv231104291P}, at low obliquity angles, and particularly at low magnetic moments where there is no interchange slingshot instability, there are major differences between the anti-parallel and parallel configurations when the pulsar is far from the propeller state. In the anti-parallel configuration, there is reconnection between the magnetic field lines from the NS and the magnetic field lines of the torus with opposite polarity. The reconnection causes more magnetic field lines to open up, therefore increasing the power of the jet more than in the isolated pulsar. Even though the difference between the two cases is more pronounced in 2D simulations, in 3D we also see this effect. As we increase obliquity, though, the difference between the anti-parallel and parallel configurations decreases, and at higher obliquity is not as influential as changing the magnetic moment or the obliquity angle. Due to the geometry, in high-obliquity pulsars reconnection between the magnetic field lines of the NS and the torus is less effective. In that sense, high-obliquity pulsars are more similar to the parallel configurations than the anti-parallel. 

In addition, high obliquity angles are geometrically less favorable for flux opening in general, as, beyond any given radius, less of the pulsar's closed flux passes through the region near the rotational equator, where it can interact with the accretion flow. This would also suppress flux opening via magnetic interchange.

These trends are clear in Figure~\ref{fig:quantities_time}, as the open magnetic fluxes --- and therefore jet powers --- for the higher-obliquity cases are lower than for the isolated pulsar case, just like for the small-$\mu$ parallel-orientation configurations with low obliquity angles.

\subsection{Perpendicular rotator}
\label{sec:perp-rot}

Figure~\ref{fig:perpendicular} shows density snapshots of our perpendicular rotator, $\chi=90^\circ$, for $\mu=10$. In the \textit{top} panel, we show three different planes. From left to right, we are showing (a) the plane perpendicular to $\mu-\Omega$, (b) the $\mu-\Omega$ plane, and (c) a plane between these two. We choose to show this particular magnetic moment, since at higher magnetic moments, the pulsar is no longer in an accretor state. Similar to a $\chi=60^\circ$ pulsar, accretion occurs in thin streams through the magnetic poles. We can also note that accretion is not limited to the $\mu-\Omega$ plane, or the plane perpendicular to the $\mu-\Omega$ plane. This means that the hot-spot (where material falls onto the NS) is not a simple ring, but has a rather complicated geometry. Indeed, unlike an aligned rotator, the accretion does not occur in a ``ring''-like manner. %

\section{Conclusions}
\label{sec:summary}
In this work, we present the first systematic exploration, via 3D GRMHD simulations, of gas interaction with rapidly rotating oblique NS magnetospheres, spanning the full range of NS field strength (from the accretor to propeller) and magnetic obliquity (from aligned to orthogonal).
For the first time for oblique NS accretion, we study the effect of disk magnetic polarity: anti-parallel and parallel polarity relative to the NS closed-zone magnetic field (able and unable to magnetically reconnect with the NS closed-zone field, respectively).

\subsection{Strong-field Accretor Regime}
\label{sec:strong-field-accretor}
Let us first consider the case of relatively strong stellar dipolar magnetic field, $\mu = 20$, shown with green lines in Figure~\ref{fig:quantities_angle}. Figure~\ref{fig:quantities_angle} shows that as the inclination angle increases, the magnetosphere pushes the gas further away (Figure~\ref{fig:quantities_angle}e), which suppresses the accretion (Figure~\ref{fig:quantities_angle}a), closes more magnetic flux (Figure~\ref{fig:quantities_angle}b), lowers jet luminosity (Figure~\ref{fig:quantities_angle}c), and weakens spindown torques (Figure~\ref{fig:quantities_angle}d). Why does this happen? One possibility is that the magnetic field strength of a dipole is twice as high at its pole as at its equator: therefore tilting the pole increases the field strength at the rotational equator in the $\mu$--$\Omega$ plane and might be able to push away the gas more efficiently. Figure~\ref{fig:quantities_angle}  (orange symbols) shows that the results are similar for $\mu = 10$.

While preparing this manuscript for publication, we became aware of a recent preprint \citep{2023arXiv231105301D} that also describes 3D GRMHD simulations of accretion onto an oblique rotator. Their simulations also show that the jet power decreases as the obliquity angle increases, therefore agreeing with our results for $\mu\sim10-20$. 

Interestingly, there is little difference between the parallel and anti-parallel torus magnetic field polarities for $10\lesssim \mu\lesssim20$: for instance, Figure~\ref{fig:quantities_angle}b--e shows that for $\mu = 20$, the green dashed and solid lines, which show parallel and anti-parallel polarities, respectively, are very close to each other. 

A mechanism that leads to such similarity of the two magnetic polarities is the \emph{interchange slingshot} instability \citep{2023arXiv231104291P}, which is a variation of the magnetic Rayleigh-Taylor instability that is activated when the magnetospheric field is strong enough to push the gas close to, or into, the propeller state. This instability, which is most vigorous when the disk and magnetospheric magnetic field lines are parallel (so they cannot reconnect), mixes the closed magnetospheric magnetic field lines into the disk, causing them to open up and power the twin polar jets.

\subsection{Weak-field Accretor Regime}
\label{sec:weak-field-accretor}

\begin{figure}[t!]

\includegraphics[scale=0.52]{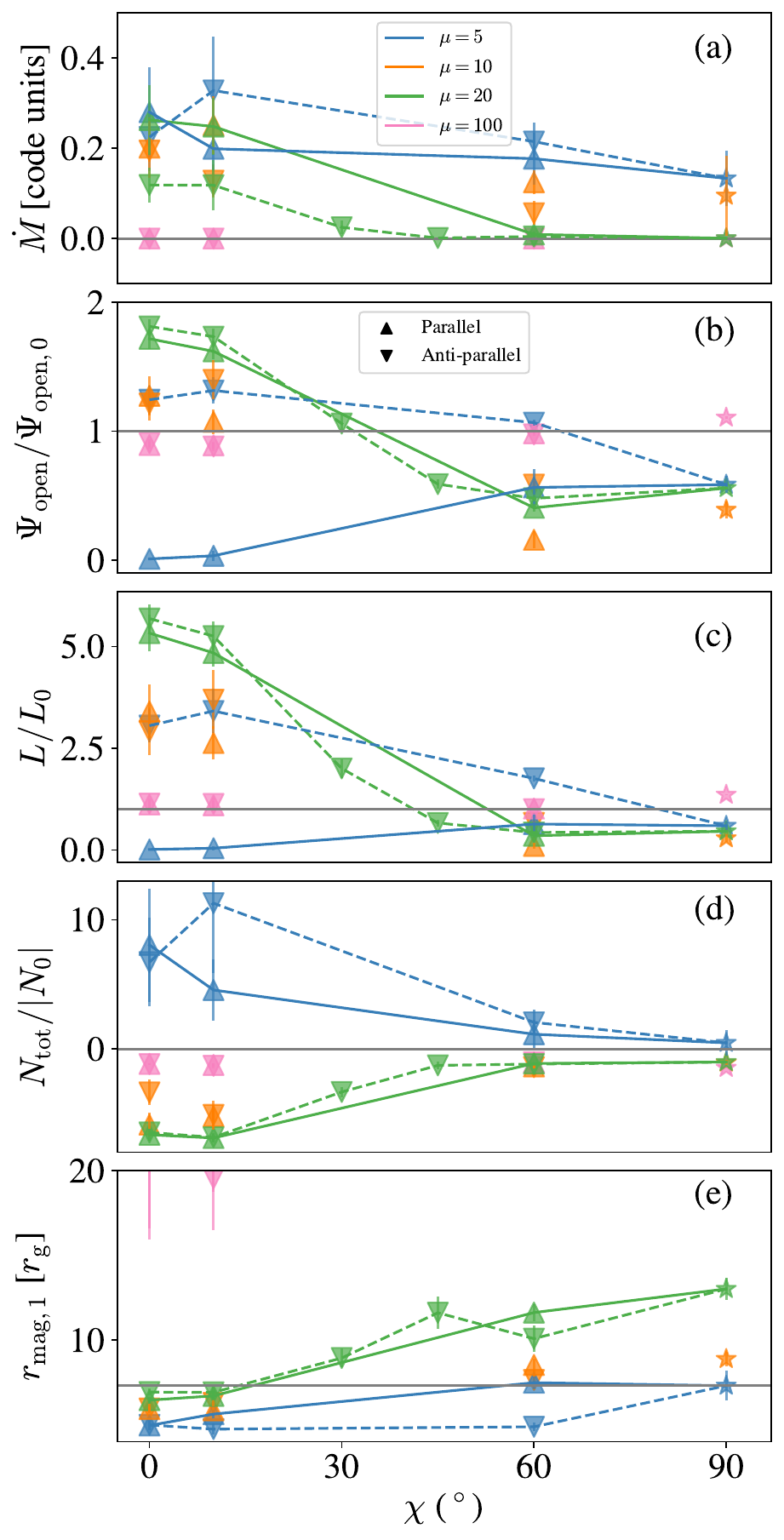}
\centering
\caption{Several quantities averaged in time, as a function of obliquity angle. The quantities are the same as in Figure~\ref{fig:quantities_time}. The grey horizontal lines represent $\Psi_{\rm open}=\Psi_{\rm open,0}$, $L=L_0$, $N_{\rm tot}=-|N_0|$, $r_{\rm mag}=\rco$.   We averaged at times when the quantities are in the steady state ($t>10{,}000\,r_{\rm g}/c$); the error corresponds to the standard deviation over that time period. We show lines that connect the values for simulations with $\mu=20$ (green), $\mu=5$ (blue) for both parallel (solid line), and anti-parallel (dashed line) configurations.}
\label{fig:quantities_angle}
\end{figure}

With the help of the interchange slingshot instability, even the parallel magnetic configuration can open magnetic field lines and launch jets for high enough $\mu$ values. What happens at weaker stellar fields? We find that the system can show more complex behavior. For example, at $\mu=5$, Figure~\ref{fig:quantities_angle} shows that the two different polarities exhibit very different behavior. In the anti-parallel configuration there is reconnection of the magnetic field lines from the torus and the magnetosphere that increases the open magnetic flux (Figure~\ref{fig:quantities_angle}b), whereas in the parallel configuration there is no reconnection, and no opening of the magnetic flux. This suppresses the jet power.  
However, as $\chi$ increases and approaches the perpendicular case, the difference between the two configurations becomes smaller.

What causes the difference between the polarities to vanish at high obliquity?
Firstly, the two polarities are equivalent for an orthogonal rotator ($\chi=90^\circ$), and one would naturally expect the polarity-dependence to disappear gradually as this limit is approached.
In addition, Figure~\ref{fig:quantities_angle}e shows that high-obliquity pulsars approach the propeller regime, $r_{\rm mag}\gtrsim\rco$. For $\mu = 5$, the transition to the propeller regime happens at $\chi\simeq60^\circ$, leading to the activation of the interchange slingshot instability that opens up the magnetic flux and therefore increases the jet power, as seen in Figure~\ref{fig:quantities_angle}c. Remarkably, this leads to opposing jet power trends with the obliquity angle for the two star--disk magnetic orientations at $\mu = 5$: whereas $L(\chi)$~decreases for the anti-parallel polarity, $L(\chi)$~increases for the parallel polarity.
We conclude that the $\chi$-dependence of the various key quantities also significantly depends on the magnetic moment and the relative magnetic polarity.

\subsection{Propeller Regime}
\label{sec:propeller}

Our simulations show that high-obliquity pulsars enter the propeller regime ($r_{\rm mag}>\rco$) at lower magnetic moments than low-obliquity or aligned pulsars. 
If we focus on low magnetic moments ($\mu=5$), we can see that in all cases, we have some non-zero mass accretion rate that decreases with increasing $\chi$ (Figure~\ref{fig:quantities_angle}a). High-obliquity pulsars accrete through a thin stream, whereas low-obliquity pulsars accrete in a ``ring''-like manner. For $\mu = 5$,  $r_{\rm mag}\lesssim \rco$ at all values of $\chi$ (Figure~\ref{fig:quantities_angle}e),  thus all pulsars we simulated for $\mu=5$ are in the accretor regime. If we increase the magnetic moment while keeping the gas reservoir constant, the difference between the low- and high-obliquity cases becomes more noticeable. The mass accretion rate decreases, and the magnetospheric radius increases as $\chi$ increases at constant $\mu$. In our simulations, the high-obliquity pulsars reach the propeller ($r_{\rm mag}>\rco$) regime between $\mu=5$ and $\mu=10$, whereas the aligned rotators change state between $\mu=20$ and $\mu=100$. 
It is interesting that sometimes pulsars in the propeller regime exhibit a non-zero mass accretion rate.

Through mapping out the parameter space, we explored the transitions between the accretor and propeller regimes. For example, for $\mu=20$,  Figure~\ref{fig:quantities_angle} shows  two transitions somewhere between $\chi\sim30^\circ$ and $\chi\sim45^\circ$: the magnetospheric radius passes through and exceeds the corotation radius, and the jet power goes from being larger than the isolated-pulsar spindown power (i.e. $L>L_0$) to being less than this value. These intermediate states occur between the full accretor and full propeller states.

\begin{acknowledgements}

This research was supported in part through the computational resources and staff contributions provided for the Quest high performance computing facility at Northwestern University which is jointly supported by the Office of the Provost, the Office for Research, and Northwestern University Information Technology.
A.M-B.\ is supported by NASA through the NASA Hubble Fellowship grant HST-HF2-51487.001-A awarded by the Space Telescope Science Institute, which is operated by the Association of Universities for Research in Astronomy, Inc., for NASA, under contract NAS5-26555.
KP, JJ, and AT acknowledge support by NASA 80NSSC21K1746 grant. AT and JJ also acknowledge support by NASA 80NSSC22K0938 and NSF AST-2009884 grants. AT was additionally supported by NSF grants AST-2107839, AST-1815304, AST-1911080, OAC-2031997, and AST-2206471. KP was supported in part by the Laboratory Directed Research and Development Program at Princeton Plasma Physics Laboratory, a national laboratory operated by Princeton University for the U.S.\ Department of Energy under Prime Contract No.\ DE-AC02-09CH11466. Support for this work was also provided by the National Aeronautics and Space Administration through Chandra Award Number TM1-22005X issued by the Chandra X-ray Center, which is operated by the Smithsonian Astrophysical Observatory for and on behalf of the National Aeronautics Space Administration under contract NAS8-03060. This research was also made possible by NSF PRAC award no. 1615281 at the Blue Waters sustained-petascale computing project and supported in part under grant no. NSF PHY-1125915. This research used resources of the Oak Ridge Leadership Computing Facility, which is a DOE Office of Science User Facility supported under Contract DE-AC05-00OR22725 via ALCC, INCITE, and Director Discretionary allocations PHY129. 
\end{acknowledgements}

\software{\textsc{harmpi} \citep{2019ascl.soft12014T}, matplotlib \citep{matplotlib},  numpy \citep{numpy}, scipy \citep{scipy}, hdf5 \citep{hdf5}, yt \citep{Turk2011}} 

\section*{Data Availability}
	
The data underlying this article will be shared on reasonable request to the corresponding author.

\bibliography{ns_bib.bib}

\end{document}